# MOHCS: Towards Mining Overlapping Highly Connected Subgraphs*

Xiahong Lin
School of Computer Science and Technology
Xidian University
Xi'an, China
xhlin@sjtu.edu.cn

Lin Gao†
School of Computer Science and Technology
Xidian University
Xi'an, China
lgao@mail.xidian.edu.cn

Kefei Chen
Department of Computer Science and Engineering
Shanghai Jiaotong University
Shanghai, China
kfchen@sjtu.edu.cn

David K. Y. Chiu
Department of Computing and Information Science
University of Guelph
Guelph, N1G 2W1, Canada
dchiu@cis.uoguelph.ca

*Abstract*—Many networks in real-life typically contain parts in which some nodes are more highly connected to each other than the other nodes of the network. The collection of such nodes are usually called clusters, communities, cohesive groups or modules. In graph terminology, it is called highly connected graph. In this paper, we first prove some properties related to highly connected graph. Based on these properties, we then redefine the highly connected subgraph which results in an algorithm that determines whether a given graph is highly connected in linear time. Then we present a computationally efficient algorithm, called MOHCS, for mining overlapping highly connected subgraphs. We have evaluated experimentally the performance of MOHCS using real and synthetic data sets from computer-generated graph and yeast protein network. Our results show that MOHCS is effective and reliable in finding overlapping highly connected subgraphs.

## I. Introduction

In a graph modeling a network, such as biological network [1], information network [2] or social network [3], a highly connected subgraph always corresponds to a cohesive set of interconnected vertices which is meaningful. For example, a dense co-expression network may represent a tight co-expression cluster [4]. The definitions of highly connected graph may vary in different works. We define a highly connected graph (or simply dense graph) as a graph whose minimum cut is no less than half of its vertex set size (and the formal definition can be found in [5]). Due to its wide application, identifying these a priori unknown building blocks is crucial to the understanding of the structural and functional properties of networks. Researchers have addressed various problem settings and have proposed numerous algorithms to achieve their goals in the past. Our review only focuses on the algorithms that are most related to our work. Among those that are most related to our work, [6] provides a definition of highly connected subgraph that is valid and useful in practice. There, the HCS algorithm is one of the most well-known clustering algorithms and has been widely used in various domains such as gene expression analysis [7] and functional module discovery [5, 8-10]. It recursively partitions the current graph into two subgraphs by removing the minimum cut until the graph is highly connected [5]. However, HCS has some shortcomings. First, HCS cannot identify overlapping highly connected subgraphs because of its nature of graph-partitioning [5]. Second, when applying the algorithm repeatedly to a large and sparse graph, HCS often cuts off one vertex in each iteration, thus having time complexity of $O(|V|^2|E|+|V|^3\log|V|)$ [5]. Third, the minimum cut algorithm is a critical step used in HCS. However, when applying HCS to a graph with numerous edges closed to quadratic, the fastest deterministic minimum cut algorithm [11] has time complexity of $O(|V|^3)$. In [5], Hu et al. proposed an algorithm called MODES, combining HCS with normalized cut, and designed a procedure to identify overlapping highly connected subgraphs. Furthermore, to mine highly connected subgraphs more effectively, several authors introduce greedy vertex deletion algorithm based on the observation that in order to produce a highly connected subgraph, the low degree vertices can be disregarded intuitively. For example Asahiro et al. [12] proposed the following greedy algorithm to find a $k$-vertex subgraph with the maximum weight: repeatedly remove a vertex with the minimum weighted-degree in the currently remaining graph, until exactly $k$ vertices are left. However, these greedy algorithms can not be used directly to our problem because of the differences in the definition and other problem settings which we will explain below.

Our motivation is to find a more efficient algorithm for mining overlapping highly connected subgraphs by reconsidering the properties of highly connected subgraph. The contributions of our work are follows:

- We give several properties and consider them in a new definition of highly connected subgraph.

* Supported by National Natural Science Foundation of China (No. 60574039) and the Project Sponsored by the Scientific Research Foundation for the Returned Overseas Chinese Scholars, State Education Ministry.
† To whom correspondence should be addressed.
E-mail: lgao@mail.xidian.edu.cn

- We present an algorithm for determining a highly connected subgraph in linear time.
- We propose an efficient algorithm MOHCS for Mining Overlapping Highly Connected Subgraphs (MOHCS).

Although MOHCS is also a greedy vertex deletion algorithm, it is applicable only after a derivation from the set of properties we have discovered. What is more, we can also identify overlapping highly connected subgraphs using a procedure in [5].

The rest of the paper is organized as follows. In section II, we introduce some notations and preliminary concepts to study some properties related to highly connected graph. In section III, the MOHCS algorithm is provided and its complexity is discussed. The refinement of the MOHCS algorithm is then presented in section IV. Section V provides a detailed experimental evaluation of MOHCS using real and synthetic data sets. Finally, we conclude our work in section VI. A preliminary version of our work could be found in [19].

## II. PROPERTIES OF HIGHLY CONNECTED GRAPH

In this section, we first introduce some notations and preliminary concepts in order to simplify our discussion. Next we give a tighter lemma on minimum cut than the one given in [6]. Then we present a relation between the minimum cut and the minimum degree which gives us new insight into highly connected graphs and derives an efficient algorithm for their identification.

### A. Notations and definitions

The notations that will be used throughout the paper are summarized in Table 1.

*Definition 1:* (Induced Subgraph [13]) Given a graph $G=(V,E)$ and a mapping $f: E \to V \times V$, an induced subgraph is a graph $G(V_s)=(V_s, E_s)$, where $V_s \subseteq V$, $E_s \subseteq E$ and $\forall v_i, v_j \in V_s$

$$e_h = (v_i, v_j) \in E_s \Leftrightarrow f(e_h) = (v_i, v_j) \in E.$$

In other words, an induced subgraph of a graph $G$ is a subset of the vertices of $V(G)$ together with all of the edges that connect them in $G$.

*Definition 2:* (Edge Cut and Edge Connectivity) Given a graph $G=(V,E)$, an edge cut is a set of edges $E_c$ such that $G'=(V, E-E_c)$ is disconnected. A minimum cut $S$ is the smallest set among all edge cuts. The edge connectivity of $G$, denoted by $k(G)$, is equal to the size of the minimum cut $|S|$. To make our following conclusions general, here we regulate that $k(G)=0$ if $G$ is disconnected, which means we needn't to remove any edge from $G$.

TABLE I. NOTATIONS USED THROUGHOUT THE PAPER

| Notations | Description |
|---|---|
| $G$ | $G=(V,E)$, an undirected graph |
| $V(G)$ | $V=\{v_1, v_2, ..., v_k\}$, the vertex set of $G$ |
| $E(G)$ | $E \subseteq V \times V$, the edge set of $G$ |
| $G(V_s)$ | the induced subgraph on $V_s$ from $G$, $V_s \subseteq V(G)$ |
| $\deg G(v)$ | the degree of the vertex $v$ in $G$ |
| $\delta(G)$ | the minimum degree of a vertex in $G$ |
| $S$ | the minimum cut of $G$ |

*Definition 3:* (Highly Connected Subgraph or simply Dense Subgraph) A graph $G=(V,E)$ with $|V(G)|>3$ vertices is called highly connected if $k(G) \geq (|V(G)|/2)$. Note that a highly connected graph with vertex size less than 4 is trivial. A highly connected subgraph is an induced subgraph $H \subseteq G$, such that $H$ is highly connected. Some subgraphs are overlapping if they have some vertices or some edges in common.

In our problem setting, we consider the simple undirected unweighted graph only. The graph does not need to be connected. In this paper, we study the problem of mining the set of highly connected subgraphs in $G$. For instance, there are two highly connected subgraphs $G_1(\{a,b,c,d,e\})$ and $G_2(\{f,g,h,i,j,k\})$ in Figure 1.

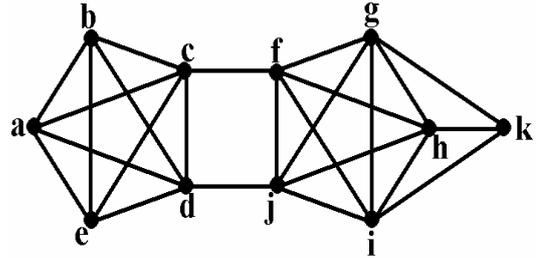

Figure 1. An example of highly connected subgraph

## B. Lemmas and Theorems

In [6], Hartuv and Shamir give a lemma as follows:

*Lemma 1:* If $S$ is a minimum cut which splits a graph $G$ into two induced subgraphs, the smaller of which, $\overline{H}$, contains $|V(\overline{H})| > 1$ vertices, then $|S| \leq |V(\overline{H})|$, with equality only if $\overline{H}$ is a clique.

Here, we first present a lemma that is similar to but tighter than Lemma 1. Although the proof procedure can be further simplified, we adopt the similar proof method used in [6] for conveniently comparing with that of Lemma 1.

*Lemma 2:* If $S$ is a minimum cut which splits a graph $G$ into two induced subgraphs, the smaller of which, $\overline{H}$, contains $|V(\overline{H})| > 1$ vertices, then $\delta(G) \leq |V(\overline{H})|$, with equality only if $\overline{H}$ is a clique.

*Proof:* When all edges are incident on a vertex with the minimum degree are removed, a disconnected graph is resulted. Therefore the edge connectivity of a graph is not greater than its minimum degree $\delta(G) \geq |S|$ [6].

Let $\deg \overline{H}(x)$ denotes the degree of vertex $x$ in $\overline{H}$, and let $\deg S(x)$ denotes the number of edges in $S$ that are incident on $x$.

Since $\delta(G)$ is the minimum degree of a vertex in $G$, for every $x \in \overline{H}$, we have

$$\deg \overline{H}(x) + \deg S(x) \geq \delta(G). \quad (1)$$

Summing over all vertices in $\overline{H}$ we get

$$\sum_{x \in \overline{H}} \deg \overline{H}(x) + \sum_{x \in \overline{H}} \deg S(x) \geq |V(\overline{H})| \delta(G), \quad (2)$$

or, equivalently

$$2|E(\overline{H})| + |S| \geq |V(\overline{H})| \delta(G). \quad (3)$$

Hence if $|V(\overline{H})| > 1$, then since

$$2\frac{|V(\overline{H})|(|V(\overline{H})|-1)}{2} \geq 2|E(\overline{H})|, \quad (4)$$

and,

$$\delta(G) \geq |S|, \quad (5)$$

we get

$$2\frac{|V(\overline{H})|(|V(\overline{H})|-1)}{2} + \delta(G) \geq |V(\overline{H})| \delta(G). \quad (6)$$

That is

$$|V(\overline{H})|(|V(\overline{H})|-1) \geq (|V(\overline{H})|-1)\delta(G). \quad (7)$$

Since $|V(\overline{H})| > 1$, then

$$\delta(G) \leq |V(\overline{H})|. \quad (8)$$

If $\delta(G) = |V(\overline{H})|$, then the inequalities (4) and (5) must hold as equalities, so

$$\delta(G) = |S|, \quad (9)$$

and

$$|E(\overline{H})| = \frac{|V(\overline{H})|(|V(\overline{H})|-1)}{2}, \quad (10)$$

which implies that $\overline{H}$ is a clique.

Since $\delta(G) \geq |S|$, Lemma 2 is tighter than Lemma 1. Because $\overline{H}$ is the smaller subgraph, Lemma 2 also implies that $\delta(G) \leq (|V(G)|/2)$. Next we will use Lemma 2 to derive two theorems. Although these theorems can also be proved by generic graph theory method [18], our proof procedures provide some evidences that it is the tighter lemma that leads to a faster algorithm in our proposal.

*Theorem 1:* If a graph $G$ is highly connected, then $\delta(G) = |S|$.

*Proof:* It is known that $\delta(G) \geq |S|$. Now suppose $\delta(G) > |S|$, the condition of equality in Lemma 2 does not hold, so we have $\delta(G) < |V(\overline{H})|$, that is $\delta(G) < (|V(G)|/2)$; because $G$ is highly connected, $\delta(G) > |S| \geq (|V(G)|/2)$, that is $\delta(G) > (|V(G)|/2)$; resulting in a contradiction. So we get $\delta(G) = |S|$.

Theorem 1 shows a relation between the minimum cut and the minimum degree that if $G$ is highly connected, the size of the minimum cut is equal to the minimum degree. Let $v$ be a vertex with minimum degree in $G$, then $(\{v\}, V \setminus \{x\})$ would be a minimum cut of $G$. In other words, we can determine edge connectivity of a highly connected graph in linear time.

*Theorem 2:* A graph $G = (V, E)$ is highly connected if and only if $\delta(G) \geq (|V(G)|/2)$.

*Proof:* ($\Rightarrow$) Since $G$ is highly connected, then $\delta(G) = |S| \geq (|V(G)|/2)$, by Theorem 1.

($\Leftarrow$) Since $\delta(G) \geq (|V(G)|/2)$; suppose $\delta(G) > |S|$, the condition of equality of Lemma 2 doesn't hold, so we have $\delta(G) < |V(\overline{H})|$, that is $\delta(G) < (|V(G)|/2)$; a contradiction. So $|S| = \delta(G) \geq (|V(G)|/2)$, $G$ is highly connected.

Based on Theorem 2, we can redefine highly connected graph as follow.

*Redefinition:* (Highly connected Graph) A graph $G = (V, E)$ with $|V(G)| > 1$ vertices is called highly connected if $\delta(G) \geq (|V(G)|/2)$.

## C. Determining Highly Connected Subgraph in Linear Time

By the initial definition of highly connected subgraph, if we want to determine whether a graph $G$ is highly connected, we should first apply the minimum cut algorithm on $G$ and then check whether the size of the minimum cut is no less than half of $|V|$. So determining highly connected subgraph has time complexity $O(|V||E| + |V|^2 \log |V|)$.

However based on our redefinition, we can simply determine $G$ by checking whether the degrees of all vertices of $G$ is no less than half of $|V|$ with linear time complexity $O(|V|)$.

## III. ALGORITHM AND COMPLEXITY

Note that our new definition of highly connected graph implies that the minimum degree of the graph must be less than half of its vertex set size, if a graph is not highly connected. And the vertex with minimum degree is the most conflictive vertex to our definition. To make the graph hold highly connected, we can then delete this vertex. Based on this observation, we directly design a greedy vertex deletion algorithm, MOHCS for mining overlapping highly connected subgraph. However, in order to describe the problem clearly, let us first consider the graph consisting of highly connected subgraphs without overlapping. We present the implementation details and refine our algorithm in section IV. Our MOHCS algorithm is outlined in Algorithm 1 and illustrated in Figure 2. Let *SDS* be the set of dense subgraphs. Line 1 initializes *SDS* as an empty set. The until loop of lines 2-9 repeatedly mines highly connected subgraphs, until $G' = \emptyset$ in which case no other highly connected subgraphs is left in $G$. The while loop of lines 4-5 repeatedly deletes the minimum degree vertex $v$ in $G'$ until it satisfies the new definition of highly connected graph or $G' = \emptyset$ in which case no highly connected subgraph exists in $G'$ when all vertices in $G'$ deleted. In lines 6-8, each highly connected subgraph is saved in *SDS* and removed from $G$.

For each iteration of MOHCS, our algorithm repeatedly removes a vertex with the minimum weighted-degree in the currently remaining graph, until a highly connected subgraph is resulted. This procedure is very similar to the algorithm proposed in [14] whose complexity is $O(|E| + |V| \log |V|)$. However our stop criterion is stricter. So the complexity of each iteration will never exceed that of [14]. In other words, as in [14], by using Fibonacci heaps [15] to hold vertices, we can get a running time of $O(|E| + |V| \log |V|)$ to identity one highly connected subgraph. After finding one highly connected subgraph, our algorithm saves the subgraph and deletes it from the remaining graph. This procedure takes a running time of $O(|E|)$ to achieve. So our MOHCS algorithm has time complexity $O(k|E| + k|V| \log |V|)$ to identify all highly connected subgraphs in $k$ iterations, where $k$ is the number of highly connected subgraphs in a graph.

---

let *SDS* be the set of dense subgraphs

let $v$ be the minimum degree vetex in $G'$

input $G$

1. $SDS = \emptyset$
2. do
3.     $G' = G$
4.     while $\deg(v) < |V(G')|/2$ and $G' \neq \emptyset$
5.        $G' = G \setminus \{v\}$
6.     if $G' \neq \emptyset$ then
7.        save $G'$ in *SDS*
8.        $G = G \setminus G'$
9. until $G' = \emptyset$

    output *SDS*

---

Algorithm 1: The MOHCS algorithm

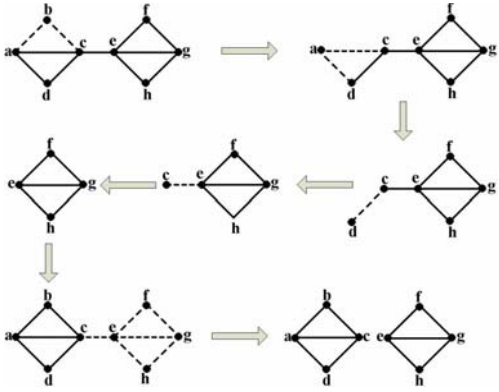

Figure 2. An example of applying MOHCS algorithm to a graph. Broken lines denotes the vertex and its neighboring edges will be deleted

Because graph is arbitrary, the performance of MOHCS would be different sharply in different graphs. When there is no highly connected subgraph in graph $G$ or $G$ itself is a highly connected graph, MOHCS terminates in time complexity of $O(|V|)$. Suppose $G'$ is a highly connected subgraph whose minimum degree is maximum among all highly connected subgraphs contained in $G$, MOHCS will mine $G'$ firstly. In reality, a highly connected subgraph with greater minimum degree always has greater vertex set size. Hence, the vertex set size of $G$ decreases quickly in each iteration of MOHCS making our algorithm having a better performance. In addition, the singleton adoption and the overlapping subgraphs identification only increases the running time of MOHCS slightly and have no influence on the time complexity of MOHCS.

## IV. REFINEMENTS OF THE MOHCS ALGORITHM

In this section, we present the implementation details and the refinements of our MOHCS algorithm. We employ the singleton adoption method proposed in [6] and the overlapping subgraphs identification method proposed in [5] with some modifications. Both original methods need several bounds to control their executions based on experimental test. However, in this study, these bounds are unnecessary due to the properties we have discovered. The complete version of MOHCS is outlined in Algorithm 2 and illustrated in Figure 4.

### A. Minimum degree vertex selection

Because of the greedy vertex deletion character of MOHCS, we use Fibonacci heaps to maintain the degrees of the vertices in the subgraph induced by $G'$. Each iteration in lines 4-5 involves identifying and removing the minimum degree vertex as well as updating the degrees of the vertices neighboring on it [14]. In fact there are always many vertices with minimum degree. Different choices of vertex for deleting lead to different results, sometimes even wrong results. Figure 3 illustrates an example that MOHCS performs incorrectly, when it collapses two highly connected subgraphs $G_1(\{a,b,c,d\})$ and $G_2(\{e,f,g,h\})$ into singletons. The reason of this case is that MOHCS chooses the minimum degree vertices alternately from two highly connected subgraphs, thus collapsing the two subgraphs. To address this problem, we import a scheme discussed as follows.

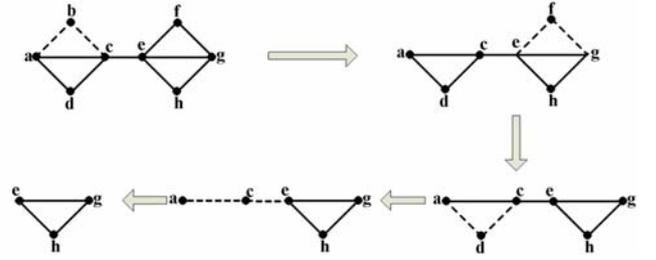

Figure 3. An example that MOHCS performances in an incorrect behavior

In a graph with many highly connected subgraphs, we aim at our MOHCS algorithm to perform in a manner such that when one vertex of a highly connected subgraph is deleted, the whole highly connected subgraph must be collapsed as well. This avoids the incorrect behavior described above. We implement this by confirming that if several vertices have the same degree, the vertex recently updated should be deleted first. Hence we use a criterion for comparison in Fibonacci heaps as bellow:

$$\forall v_i, v_j, key(v_i) < key(v_j), \text{if } \deg G(v_i) < \deg G(v_j)$$

or $\deg G(v_i) = \deg G(v_j)$ and $v_i$ is more recently updated than $v_j$.

### B. Singleton adoption

Our MOHCS algorithm may also leave some vertices as uncluttered singletons. For example, when applying MOHCS in the graph $G$ in Figure 4, the vertex $i$ is left as singleton. Hence after the execution of MOHCS, we need to check whether some singletons can be adopted by some highly connected subgraphs. Our singleton adoption is derived from the method proposed in [6]. Our proposed modification procedure is described as follows.

Let $S$ denotes the singleton set. One singleton in $S$ may fit into many highly connected subgraphs, taking into account the existence of overlapping subgraphs. For each highly connected subgraph $G'$, we decreasingly sort all singletons in $S$ by the number of neighbors they have in $G'$. Then we check every singleton in order. For each singleton $v$, if its join keeps $G'$ still highly connected (using our algorithm for determining highly connected graph), $v$ is added to $G'$. The process is repeated until there is one singleton that could not be fitted into $G'$.

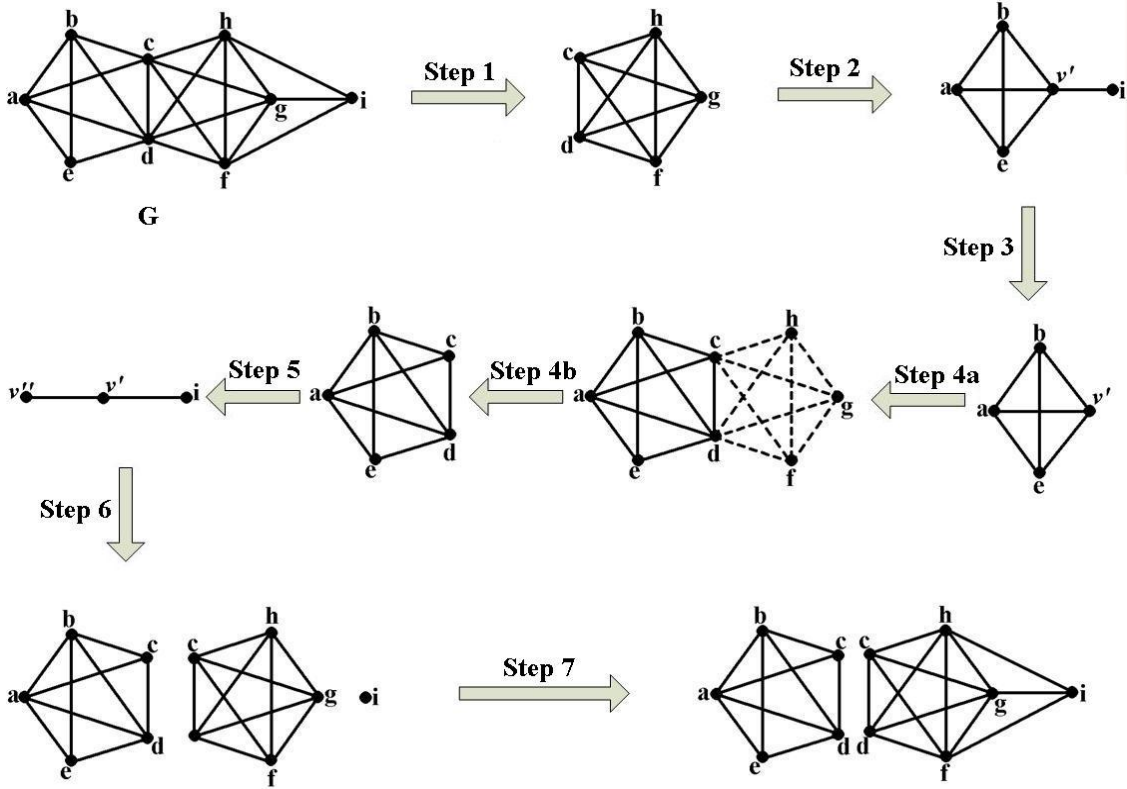

Figure 4. An example of applying the complete version of MOHCS algorithm to a graph

---

let *SDS* be the set of dense subgraphs

let *v* be the minimum degree vetex in $G'$

input *G*

1. $SDS = \varnothing$
2. do
3.     $G' = G$
4.     while $\deg(v) < |V(G')|/2$ and $G' \neq \varnothing$
5.        $G' = G' \setminus \{v\}$
6.     if $G' \neq \varnothing$ then
7.        save $G'$ in *SDS*
8.        condense $G'$ into a vertex $v'$
8.        $G = G \setminus G' \cup \{v'\}$
9. until $G' = \varnothing$
10. Overlapping subgraphs identification
11. Singleton adoption

output *SDS*

---

Algorithm 2: The complete version of MOHCS algorithm

*Overlapping subgraphs identification*

Because highly connected subgraphs may overlap with each other, if we simply remove one of these subgraphs directly, the others will be destroyed for existing vertices and edges that are in common. Considering this problem, we employed a method proposed in [5]. We restate the procedure with slight modification as follow.

When mining a highly connected subgraph $G'$ from the graph $G$ (lines 6-8), we condense $G'$ into a vertex $v'$. And any vertex $v$ in $G \setminus G'$ has an edge with $v'$ if $v$ links to at least one vertex in $G'$. In later iterations of the until loop (lines 2-9), if $v'$ is contained in a newly mined highly connected subgraph $G''$, we restore $v'$ to the original subgraph $G'$, and check all vertices of $G'$ whether they can be adopted to $G''$ by using the singleton adoption method mentioned above.

C. *The complete version of MOHCS*

So far we have developed the complete version of MOHCS presented in Algorithm 2 and illustrated in Figure 4, combining with the implementation details and the refinements described above. In Figure 4, The graph $G$ consists of two highly connected subgraphs $G_1(\{a,b,c,d,e\})$ and $G_2(\{c,d,f,g,h,i\})$. $G_1$ and $G_2$ are overlapping with each other. In Step 1, we mine $G(\{c,d,f,g,h\})$ first. In Step 2, we condense $G(\{c,d,f,g,h\})$ into a vertex $v'$. In Step 3, we mine $G(\{a,b,v',e\})$. In Step 4a, we restore $v'$ to $G(\{c,d,f,g,h\})$. In Step 4b, After adopting the vertices $c$ and $d$, $G(\{a,b,v',e\})$ becomes $G_1(\{a,b,c,d,e\})$. In Step 5, we condense $G_1(\{a,b,c,d,e\})$ into vertex $v''$. In Step 6, we leave vertex $i$ as singleton. Finally in Step 7, $G(\{c,d,f,g,h\})$ adopts the vertex $i$ and becomes $G_2(\{c,d,f,g,h,i\})$.

## V. EXPERIMENTAL EVALUATION

In this section, we first compare the algorithm MOHCS with HCS on computer-generated graph. Since both MOHCS and HCS do not apply to graphs that are weighted, we consider only those cases when they are unweighted. The result shows that our algorithm performs more reliably than the HCS algorithm. Then we apply the MOHCS algorithm to yeast protein interaction network and find out more than 1188 modules larger than three. We also verify that the size of graph decreases quickly as iterations of MOHCS.

### A. Algorithm comparison on computer- generated graphs

To compare the performance of MOHCS with HCS [6], we apply them on computer-generated graphs. Because HCS can not discover overlapping subgraphs, it is clear that our algorithm will outperform HCS in this case. Thus we just compared them on graphs without overlapping subgraphs. The construct method is based on the generic random graph model [16]. We construct a generated graph contains $k$ highly connected subgraphs, each of which has $n$ vertices. Every vertex belongs to one and only one subgraph, which ensures that different subgraphs will not have vertices or edges in common, that is, all subgraphs are not overlapping. Thus, each constructed graph has $N = nk$ vertices in total. For each vertex pair $(i, j)$, we realize an edge with fixed probability $p$ if $i$ and $j$ are in the same subgraph, and $q$ if $i$ and $j$ are in different subgraphs. Since there are $n^2$ vertex pairs between two subgraphs and $n(n-1)/2$ vertex pairs in a subgraph. By Definition 3, we know that the mathematical expectation of edge number between two subgraphs should be less than half of the vertex number of these two subgraphs, that is, $2n/2$. By our new definition, we know that the mathematical expectation of edge number in a subgraph should not be less than $n^2/4$. Then we have

$$n^2 q < n, \quad (11)$$

and,

$$\frac{n(n-1)}{2} p \geq \frac{n^2}{4}. \quad (12)$$

That is,

$$p \geq \frac{n}{2(n-1)}, \quad (13)$$

and,

$$q < \frac{1}{n}. \quad (14)$$

This method generates many graphs with known cluster structure. We examine the clustering results of MOHCS and HCS on these graphs. We find that most of highly connected subgraphs mined by HCS are also contained in that of MOHCS, except for the case that there are some subgraphs with the same density but with different vertex set. To illustrate the result more clearly, Figure 5 presents one of test graphs represented as matrix and the clustering results of MOHCS and HCS. There are ten highly connected subgraphs in the test graph. MOHCS finds out all ten subgraphs, and HCS finds only three subgraphs. Counted from the left bottom to the right top, there are some uncluttered singletons in the 1st, 5th, 6th, 8th and 10th subgraph, which can be identified by applying singleton adoption.

### B. Mining modules from yeast protein interaction network

We apply our MOHCS algorithm on yeast protein interaction network to mine modules. The experimental data is baker's yeast protein interaction network downloaded from the DIP database (version Score20070219). The network includes 4966 yeast proteins and 17530 interactions. In order to identify functional modules from the network, we express the network of proteins connected by interaction as a network of connected interaction [17]. The procedure takes a graph $G$, consisting of edges connecting vertices, and produces its associated line graph $L(G)$ in which edges now represent vertices and vertices represent edges [18]. After converted by this procedure, the graph consists of 17530 vertices and 439685 edges. MOHCS identified 1188 simple modules with size larger than three. Eleven of them have size larger than one hundred (see Figure 6). The largest module contains 283 interactions. The size of remaining graph after each iteration of MOHCS is shown in Figure 7, which verifies that the size decreases very quickly.

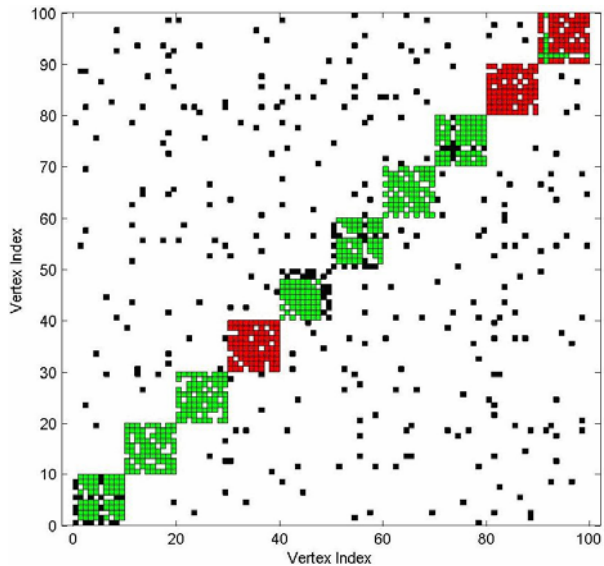

Figure 5.  A randomly cluster test matrix for $k=10$, $n=10$, $p=90\%$, $q=3\%$. Dots indicate nonzero entries. Red dots represent the clusters mined both by MOHCS and HCS. Green dots represent the clusters mined only by MOHCS

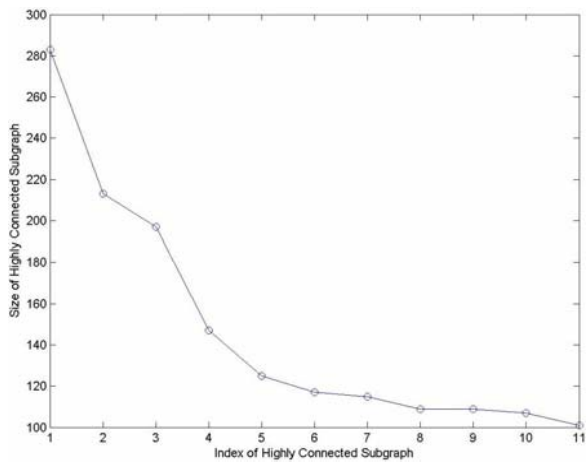

Figure 6.  Size of eleven largest subgraphs

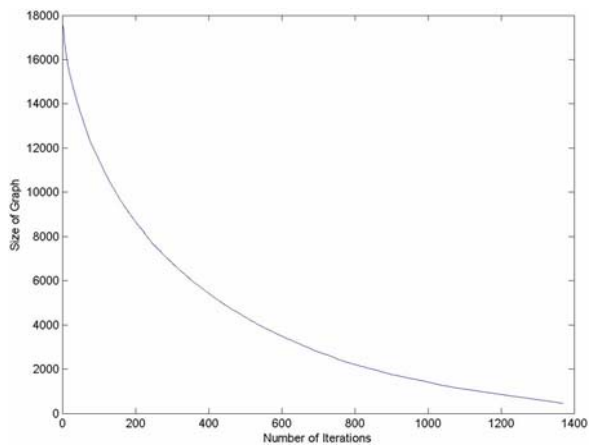

Figure 7.  Size of remaining graph as iteration of MOHCS

## VI. CONCLUSION

In this paper, we study some properties related to highly connected graph based on graph theoretic techniques, revealing a relationship between the minimum cut and the minimum degree in highly connected graphs. We also give a new definition of highly connected graph considering these properties. Further, we provide a method for determining whether a graph is highly connected. We propose an efficient algorithm MOHCS to mine overlapping highly connected subgraphs, based on this redefinition of highly connected graph. As mentioned above, different choices of vertex for deleting can lead to different or incorrect results. The previously developed greedy vertex deletion algorithms do not consider this case. Here, we present a scheme to confirm that our MOHCS chooses the correct vertex in the set of vertices with minimum degree. We employed two other methods separately proposed in [5] and [6] with modifications. The modified complete version of MOHCS is then presented. Finally, we analyze the running time of MOHCS and apply it to computer-generated graph and yeast protein interaction network. The experimental results show that the MOHCS algorithm outperforms the HCS algorithm both in computer-generated graph and yeast protein network.

### ACKNOWLEDGEMENTS

The authors would like to thank Runming Lu for his helpful remarks. We also want to thank the funding agencies for their financial support.